\documentclass[aps,prl,twocolumn,groupedaddress]{revtex4}

\usepackage[dvips]{graphicx}
\usepackage{latexsym}
\usepackage{amssymb}

\begin{document}
\bibliographystyle{nature}

\title{Kondo physics in tunable semiconductor nanowire quantum dots}
\author{Thomas Sand Jespersen,$^{1\ast}$ Martin Aagesen$^{1}$, Claus S\o rensen$^{1}$, Poul Erik Lindelof$^{1}$, Jesper Nyg\aa rd$^{1}$\\
\normalsize{$^{1}$Nano-Science Center, Niels Bohr Institute,}\\
\normalsize{University of Copenhagen, Universitetsparken 5, DK-2100}\\
\normalsize{Copenhagen, Denmark}\\
\normalsize{$^\ast$To whom correspondence should be addressed;
E-mail:  tsand@fys.ku.dk.} }
\date{\today}
\begin{abstract}
We have observed the Kondo effect in strongly coupled semiconducting
nanowire quantum dots. The devices are made from indium arsenide
nanowires, grown by molecular beam epitaxy, and contacted by
titanium leads. The device transparency can be tuned by changing the
potential on a gate electrode, and for increasing transparencies the
effects dominating the transport changes from Coulomb Blockade to
Universal Conductance Fluctuations with Kondo physics appearing in
the intermediate region.
\end{abstract}
\maketitle

As building blocks for quantum devices, semiconducting nanowires
offer an appealing combination of properties not found in other
systems. These properties include a wide range of material
parameters of the nanowire itself (e.g.\ the vanishing nuclear spin
of Si-nanowires or the high $g$-factor and strong spin-orbit
coupling in InAs) and the the wire geometry offering a wide choice
of contact materials (e.g.\ superconductors or magnetic materials).
Furthermore, nanowires offer the possibility of changing the crystal
composition along the wire to facilitate e.g.\ tailored optical
properties. In order to exploit these properties it is important to
investigate basic quantum dot (QD) phenomena in nanowire devices and
already, single electron transport has been demonstrated in weakly
coupled nanowire quantum dots in the Coulomb Blockade (CB)
regime\cite{Defranceschi:2003,Bjork:2004,Zhong:2005} and Universal
Conductance Fluctuations (UCF), Weak Localization and
Anti-Localization\cite{Hansen:2005} have been observed in low
resistance nanowire devices. As a key example of a many-body effect
in solid state physics, the Kondo effect in quantum dots has
received considerable theoretical and experimental attention, but
has sofar remained elusive in nanowires partly because of the lack
of barrier tunability. We here report on semiconducting nanowire
devices with electrostatically tunable tunnel barriers, and the
observation of the Kondo effect in the regime where the device acts
as a strongly coupled quantum dot. We find that the Lande $g$-factor
is significantly different from the bulk value for InAs due to the
strong confinement of the wire. Moreover, indications of
non-equilibrium Kondo effects are observed.\\
\begin{figure}
        \centering
        \includegraphics[width=8.5cm]{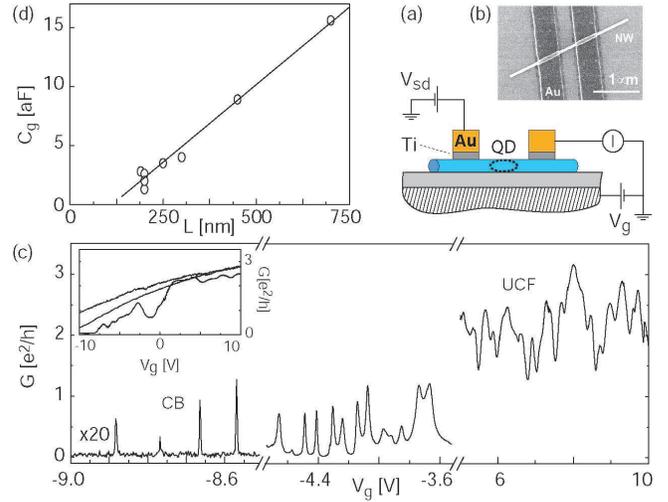}
        \caption{(color online) (a) Side-view schematic of a nanowire quantum dot and the electrical setup.
        (b) A scanning electron micrograph showing a nanowire device with
$300 \,\mathrm{nm}$ electrode separation. (c) Linear conductance
measurements $G(V_g)$ at $T=0.3\,\mathrm K$ and for temperatures $T=
270\,\mathrm{K}$, $130\,\mathrm{K}$, $15\,\mathrm{K}$ (inset, top,
middle, bottom curve). At $0.3\,\mathrm{K}$ and $V_g
> 2\, \mathrm V$ the contacts are open and UCF are observed.
For $-6\, \mathrm V < V_g < -1\, \mathrm V$ Kondo correlations are
observed and for $V_g < -7V$ the device is dominated by CB. (d)
QD-backgate capacitance $C_g$ as a function of contact separation
$L$.}
\end{figure}%
Our devices are based on semiconducting InAs-nanowires grown by
molecular beam epitaxy (MBE) from \emph{in-situ} deposited
Au-catalyst particles. The wire diameters are $\sim 70 \,
\mathrm{nm}$ and the lengths $\mathrm{2-5}\,\mu \mathrm m$. These
are the first measurements on MBE-grown InAs-nanowires and we expect
that this well-established technique for growing ultra high quality
III-V materials will produce wires of higher crystal quality and
lower impurity concentrations than wires grown by other methods,
thus resulting in cleaner systems for low temperature transport.
Details on the growth and characterization of the wires will be
published elsewhere \cite{aagesen:2006}. The wires are transferred
from the growth substrate to a highly doped Si-substrate with a
$500\, \mathrm{nm}$ insulating $\mathrm{SiO}_2$ cap-layer by gently
pressing the two wafers together. By optical microscopy the wires
are located with respect to predefined alignment marks and contacts
are defined by E-beam lithography and subsequent evaporation of
Ti/Au ($10\, \mathrm{nm}$/$60\,\mathrm{nm}$). To obtain clean
metal-nanowire interfaces the devices are treated by a 10 sec.\
oxygen plasma etch followed by a $5\,\mathrm{sec.}$\ wet etch in
buffered hydrofluoric acid prior to metal evaporation. Omitting
these last steps results in high resistance devices. Figure 1(a)
shows a device schematic and Fig.\ 1(b) a scanning electron
micrograph of a typical device. The substrate acts as a back-gate
electrode and for a given temperature $T$ and external perpendicular
magnetic field $B$ the differential conductance $dI/dV_{sd}$ is
measured as a function of the applied source-drain voltage $V_{sd}$
and back-gate
potential $V_g$ using standard lock-in methods.\\
Fig.\ 1(c) shows measurements of the linear response conductance $G$
as a function of $V_g$ at different temperatures for a device with
$L=200 \, \mathrm{nm}$ electrode separation. At room temperature
(inset) the conductance varies monotonically from $0.9\,e^2/h$ at
$V_g=-10\,\mathrm V$ to $2.8\,e^2/h$ at $V_g=10\,\mathrm V$ thereby
identifying the carriers as \emph n-type (here $e$ is the electron
charge and $h$ Plancks constant). At lower temperatures the slope of
the $G(V_g)$-trace increases but the conductance at $V_g =
10\,\mathrm V$ remains effectively unchanged. This indicates a low
barrier at the contacts at $V_g = 10\,\mathrm V$ increasing with
smaller gate-voltages. At $0.3\,\mathrm{K}$ (main panel), the
behavior is as follows: For $V_g \lesssim -7.5\,\mathrm V $ the
device behaves as a quantum dot in the CB regime with large tunnel
barriers between the leads and the dot with transport exhibiting
sharp peaks separated by regions of zero
conductance\cite{Kouwenhoven:1997}. For $V_g \gtrsim 2\,\mathrm V$
reproducible oscillations due to UCF are observed indicating low
barriers between the leads and the QD. In the intermediate region
the CB peaks are broadened due to a stronger coupling of the QD to
the leads and Kondo physics emerges as discussed below.
The periodic pattern of CB-peaks shows that the device behaves as a
single dot rather than multiple dots in
series\cite{Defranceschi:2003}. The average peak separations $\Delta
V_g \approx 80\, \mathrm{mV}$ determine the capacitance from the dot
to the back gate $C_g = e/\Delta V_g \sim 2\,\mathrm{aF}$ and the
linear dependence of $C_g$ on the contact separation (Fig. 1.d)
shows that the QD is defined by barriers at the contacts rather than
defects along the wire which would result in random dot size
fluctuations. The curve extrapolates to zero at $L_0 \approx 100 \,
\mathrm{nm}$ suggesting a significant depleted region at the
contacts. For metal contacts to bulk InAs it has been shown that the
Fermi level pins in the conduction band (resulting in good
electrical contact\cite{Bhargava:1997}), and assuming a similar
behavior for InAs nanowires we believe that the QD is formed in the
middle of the wire segment between the contacts due to the
nonuniform electrical potential produced by the backgate and the
grounded contacts.\\
\begin{figure}%
        \centering
        \includegraphics[width=8.5cm]{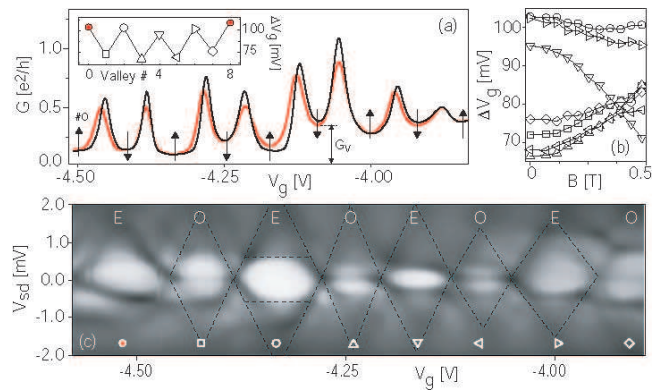}
        \caption{(color online) (a) $G$ vs.\ $V_g$ for temperatures $T=300\, \mathrm{mK}$ (black) and $T\approx 800\, \mathrm{mK}$ (red)
        showing overlapping CB-peaks. Alternating high/low
        valley conductance with corresponding decreasing/increasing behavior for increasing
        sample temperature (arrows) is observed.
          The peak spacings $\Delta
         V_g$ (inset) and their magnetic field dependence [panel (b)] follow a similar even-odd pattern.
         (d) Corresponding stability diagram (white[black] corresponds to $0e^2/h$ [$1.5e^2/h$]) showing
faintly visible CB-diamonds indicated by black lines and marked E/O
according to even/odd electron occupation number $N$. A zero-bias
Kondo resonance is observed through each odd-$N$ CB diamond. The
truncation of diamonds (horizontal features) indicate the onset of
inelastic co-tunneling.}
\end{figure}%
We now focus on the intermediate gate-region of Fig.\ 1(c). Black
and red traces of Fig.\ 2(a) show $G(V_g)$ for $T= 300 \mathrm{mK}$
and $T \approx 800 \,\mathrm{mK}$, respectively and a series of
broadened CB peaks are observed with each valley corresponding to a
fixed number of electrons ($N$) on the dot. The non-zero valley
conductance $G_v$ indicate a significant contribution to the
conductance from elastic co-tunneling processes exhibits an
alternating pattern where high-$G_v$ valleys are followed by valleys
of lower $G_v$ and vice versa. The even-odd pattern repeats in the
addition energies (peak separations $\Delta V_g$, see inset)
suggesting a twofold degeneracy of the dot levels. The electron spin
is identified as the origin of this degeneracy by measuring the
evolution of the peaks in a magnetic field as shown in Fig.\ 2(b).
Valleys corresponding to odd(even) $N$ are expected to widen(shrink)
by the Zeeman splitting $g\mu_BB$\cite{Cobden:1998} ($\mu_B$ is the
Bohr magneton) and this pairing behavior is readily observed. Figure
2(c) shows $dI/dV_{sd}$ as a function of $V_{sd}$ and $V_g$
(stability diagram). A pattern of low conductance CB-diamonds
(dashed lines), is visible, and the diamonds are numbered E/O
according to even/odd $N$ as determined from the magnetic field
dependency. Pronounced conductance ridges at zero source-drain bias
appear in every odd-$N$ diamond. These observations are consistent
with the Kondo effect: Whenever the dot holds an unpaired electron
(odd $N$), its spin is screened by the conduction electrons in the
contacts through multiple spin-flip tunnel events giving rise to a
transport resonance at zero
bias\cite{Glazman:2005,GoldhaberNature:1998}. When an even number of
electrons reside on the dot the net spin is zero and no Kondo
screening takes place.\\
\begin{figure}%
        \centering
        \includegraphics[width=6.5cm]{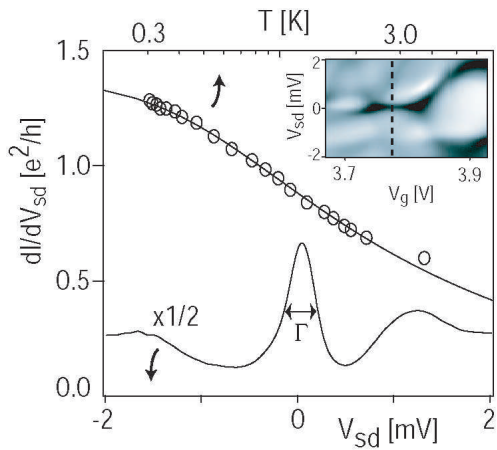}
        \caption{Top curve: temperature dependence of
the peak conductance for a particularly strong Kondo ridge. The
solid line is a fit to the formula given in the text. The inset
shows the stability diagram ($T = 300\,\mathrm{mK}$) and the lower
curve shows $dI/dV_{sd}$ vs. $V_{sd}$ through the middle of the
ridge (scaled by a factor of $1/2$).}
\end{figure}%
\begin{figure}%
        \centering
        \includegraphics[width=7.5cm]{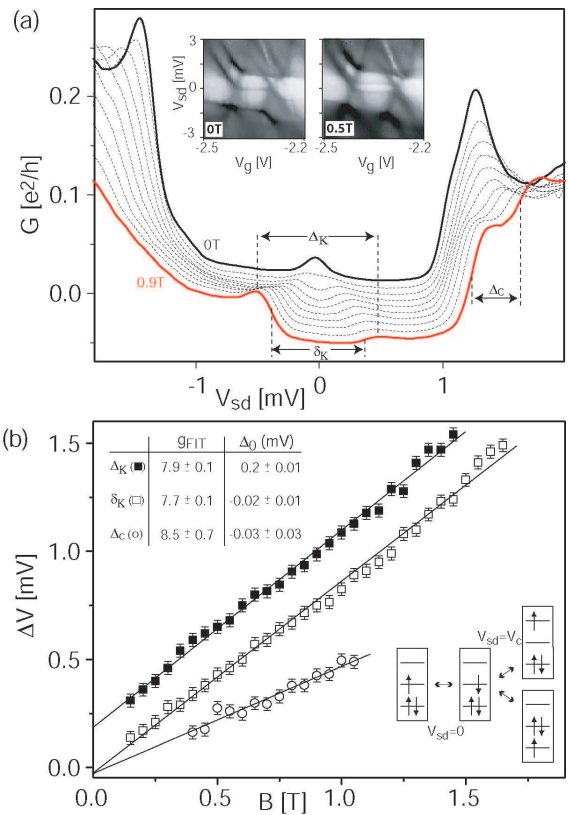}
        \caption{(color online) (a) Stability diagrams at $B=0\,\mathrm T$ (left) and $B=0.5\,\mathrm T$ of a odd-$N$
        diamond with a Kondo resonance.
        Main panel shows traces through the middle of the diamond in magnetic fields
        of $0\,\mathrm T,0.1\, \mathrm T,\dots,0.9\,\mathrm T$ (each off-set by
        $-0.0075\,e^2/h$). The peak at $V_{sd} \approx 1.25 \, \mathrm{mV}$ indicates the
        on-set of inelastic cotunneling. (b) The separation of the peaks in magnetic
        fields as indicated in (a). The upper inset shows the g-factor and extrapolated splitting $\Delta_0$
        at $B=0 \,\mathrm T$ for the linear fits.
        (Right inset) Schematic diagram of cotunneling processes for zero source-drain bias (left) giving the
        usual Kondo effect, and at finite bias (right) yielding Kondo enhanced cotunneling peaks out of equilibrium.
        All states have total spin $S=1/2$.}
\end{figure}%
In addition to the zero-bias resonances in the odd $N$ diamonds the
Kondo effect is distinguishable by unique magnetic field and
temperature dependencies. The arrows in Fig.\ 2(a) indicate the
temperature dependence of the valley conductance: In the absence of
Kondo correlations (even $N$) $G_v$ increases upon heating of the
sample as for usual CB \cite{Kouwenhoven:1997}, but in odd-$N$
valleys $G_v$ decreases. The temperature dependence of a Kondo peak
is described by the interpolation-function $G(T) =
G_0\big(T_K'^2/(T^2 + T_K'^2)\big )^s$. Here $T_K' = T_K/(2^{1/s}
-1)^{1/2}$ and $s = 0.22$ expected for a spin-half
system\cite{GoldhaberGordon:1998} and $T_K$ is the so-called Kondo
temperature. The inset to Fig.\ 3 shows a stability diagram measured
in a different cool-down of the same device. A particularly strong
Kondo resonance is observed and Fig.\ 3 shows $dI/dV_{sd}$ vs.\
$V_{sd}$ through the middle of the ridge at $300\, \mathrm{mK}$
(lower curve) and the valley conductance for different temperatures
(upper curve). The solid line shows a fit to the formula above and
the agreement is excellent yielding $T_K = \,2.1\mathrm{K}$ and $s =
0.22$ supporting the Kondo nature of the ridge. The Kondo
temperature can also be estimated from the full width at half
maximum $\Gamma \approx 2 k_B T_k /e$ of the Kondo peak and for the
data in Fig.\ 3 we find $\Gamma = 0.34 \,\mathrm{mV}$ and thus $T_K
= \Gamma e /2k_B = 2.2\,\mathrm K$ in good agreement with the
estimate above.\\
In a magnetic field the spin-degenerate states of the quantum dot
are split by the Zeeman splitting $g\mu_BB$, which for Kondo
resonances leads to peaks at $V_{sd} = \pm g\mu_BB/e$, independent
of gate voltage\cite{Glazman:2005}. The insets to Fig.\ 4(a) show
stability diagrams of a Kondo ridge measured at zero field (left
inset) and at $B = 0.5\,\mathrm{T}$ (right). A gate independent
splitting of the ridge is observed and the vertical traces through
the middle of the diamond for magnetic fields of $0 \mathrm{-} 0.9
\,\mathrm T$ (Fig.\ 4(a)) shows the evolution of the splitting.
Different methods for determining $g$ from the splitting of the
Kondo peak have been suggested in the literature. Refs.
\cite{Cronenwett:1998,Kogan:2004,Hewson:2006} suggest using the
distance $\Delta_K$ between the peaks in $dI/dV$, however, for the
weak coupling regime theoretical work has suggested the use of the
distance $\delta_K$ between peaks in $d^2I/dV^2$ (steepest points in
$dI/dV_{sd}$)\cite{Paaske:2004}. Figure 4(b) shows $\delta_K$ (open
squares) and $\Delta_K$ (solid squares) extracted from the data of
Fig.\ 4(a) and measurements at higher fields. Both $\Delta_K$ and
$\delta_K$ show a clear linear dependence, and the parameters of the
linear fits are shown in the inset with $g$ calculated from the
slope and $\Delta_0$ the extrapolated splitting at $B=0\, \mathrm
T$. Both methods gives $g \approx 8$. The considerable downshift of
the $g$-factor with respect to that of of bulk InAs ($|g_{bulk}| =
15$) reflects the confinement of the nanowire geometry and agrees
with measurements of $g$ in InAs nanowire QD's in the
CB-regime\cite{Bjork:2005}.\\
At a critical bias voltage $V_{sd} = V_c$ given by the excitation
energy for the dot, inelastic cotunneling processes which leave the
dot in an excited state set in, giving rise to horizontal ridges at
finite bias within the Coulomb diamonds (Fig.\ 2(c) and Fig.\ 4).
Figure 4(a) shows a sharp peak at zero field appearing at this onset
($V_c \approx 1.25\,\mathrm{mV}$), in addition to the zero bias
Kondo peak (odd $N$). In a magnetic field the doublet excited state
splits by $\Delta_c = g \mu_B B$, i.e.\ half that of the Kondo peak,
allowing for an independent estimate of $g$. The splitting is
readily observed in Fig.\ 4(a) and the linear dependence of
$\Delta_c$ vs.\ $B$ (Fig.\ 4(b)) gives a $g$-factor of $8.5 \pm 0.7$
consistent with that measured from the Kondo peak.\\
We note that non-equilibrium population of the excited states at
finite bias\cite{Paaske:2004} may result in a broad cusp at the
cotunneling on-set rather than a simple finite bias cotunneling
step. This mechanism, however, cannot alone account for the peak in
Fig.\ 4(a), being considerably narrower than the threshold bias and
we propose that this peak is a signature of a Kondo resonance
existing out of equilibrium. For carbon nanotubes, a detailed
quantitative analysis recently showed that for an even $N$ quantum
dot inelastic cotunneling processes can result in a non-equilibrium
singlet-triplet Kondo effect, accompanying transitions between the
singlet ground state ($S=0$) and an excited triplet state
($S=1$)\cite{Paaske:2006}. The Kondo correlations are in this case
indicated by peaks at the cotunneling onset being narrower than the
threshold bias. We see similar, sharp peaks in our devices in even
$N$ diamonds, however, for the odd $N$ example in Fig.\ 4a, the
effect coexists with the zero-bias Kondo effect and the cotunneling
peak is intensified by (spin-flipping) transitions promoting the dot
into excited states with the same total spin $S=1/2$ as the ground
state\footnote{For a dot with nearly equidistantly spaced levels
such as the region analyzed in Fig.\ 2(a) (insert), two different
excited orbital states can contribute to this resonance, cf.\ the
schematic in Fig.\ 4(b)}. No theoretical models have been solved to
allow for a quantitative description of this scenario, but
presumably, the situation is readily achieved in other Kondo dot
systems and it deserves theoretical attention.\\
The data illustrates that MBE grown InAs nanowires can constitute
highly tunable quantum dots with a rich behavior, including strongly
correlated electronic states and Kondo physics. Data from two
devices is presented, however, the results are not unique to these:
of the 9 devices investigated so far, the Kondo effect was observed
in 4; the remaining 5 exhibiting only CB. Thus, given the recent
advances in making superconducting contacts to
nanowires\cite{Doh:2005} the present results show the promise of
using nanowires for studying in a semiconductor system the interplay
between two of the most pronounced many body effects in solid state
physics: superconductivity and the Kondo effect. This subject has
been experimentally addressable only in carbon nanotube based
systems\cite{Buitelaar:2002}.\\
For helpful discussions we acknowledge K. Grove-Rasmussen, J.
Paaske, C. Marcus, K. Flensberg, C. Flint, and E. Johnson.

\end{document}